\documentclass[aps,prl,twocolumn,superscriptaddress,showpacs,floatfix]{revtex4-1}
\usepackage{amsmath,amssymb,amsfonts,dsfont, float,graphics,epsfig,epstopdf,color,verbatim,tabularx,bm,multirow}
\usepackage{wasysym}
\usepackage[utf8]{inputenc}
\usepackage[T1]{fontenc}
\usepackage{xcolor}

\IfFileExists{newtxtext.sty}
   {\usepackage{newtxtext,newtxmath}}
   {\IfFileExists{stix.sty}
      {\usepackage{stix}}
      {\IfFileExists{mathptmx.sty}
      {\usepackage{mathptmx}}{} } }

\usepackage{hyperref}
\hypersetup{
	colorlinks = true,
	linkcolor = [rgb]{0.70,0.13,0.13},
	citecolor = [rgb]{0.13,0.55,0.13},
	urlcolor = [rgb]{0.25,0.41,0.88}}

\newcommand{\dd}{\mathrm{d}}

\DeclareSymbolFont{sfletters}{OML}{cmbrm}{m}{it}
\DeclareMathSymbol{\sfeps}{\mathord}{sfletters}{"22}

\newcommand{\mb}[1]{ {\mathbf{#1}}}

\newcommand{\eqnref}[1]{Eq.\,\eqref{#1}}

\newcommand{\Refcite}[1]{Ref.\,[\onlinecite{#1}]}

\graphicspath{{Figures/}}

\def\Z{\mathbb{Z}}

\begin{document}
\title{Scaling of disorder operator at $(2+1)d$ U(1) quantum criticality}
\author{Yan-Cheng Wang}
\affiliation{School of Materials Science and Physics, China University of Mining and Technology, Xuzhou 221116, China}

\author{Meng Cheng}
\email{m.cheng@yale.edu}
\affiliation{Department of Physics, Yale University, New Haven, CT 06520-8120, U.S.A}

\author{Zi Yang Meng}
\email{zymeng@hku.hk}
\affiliation{Department of Physics and HKU-UCAS Joint Institute of Theoretical and Computational Physics, The University of Hong Kong, Pokfulam Road, Hong Kong SAR, China}

\begin{abstract}
We study disorder operator, defined as a symmetry transformation applied to a finite region, across a continuous quantum phase transition in $(2+1)d$. We show analytically that at a conformally-invariant critical point with U(1) symmetry, the disorder operator with a small U(1) rotation angle defined on a rectangle region exhibits power-law scaling with the perimeter of the rectangle. The exponent is proportional to the current central charge of the critical theory. Such a universal scaling behavior is due to the sharp corners of the region and we further obtain a general formula for the exponent when the corner is nearly smooth. To probe the full parameter regime, we carry out systematic computation of the U(1) disorder parameter in the square lattice Bose-Hubbard model across the superfluid-insulator transition with large-scale quantum Monte Carlo simulations, and confirm the presence of the universal corner correction. The exponent of the corner term determined from numerical simulations agrees well with the analytical predictions.
\end{abstract}
\date{\today}
\maketitle

{\it{Introduction.-}}
Spontaneous symmetry breaking is a fundamental phenomenon in nature. Symmetry-preserving states without ordering are often called ``disordered''. While they might appear featureless at first sight, recent advances in the classification of quantum states~\cite{WenXG17} have revealed a rich structure underlying quantum disordered phases, as condensation of extended objects, such as symmetry domain walls or field lines of emergent gauge field~\cite{WenXG17,WenXG19}. Such hidden structures completely escape the grasp of local measurement, and non-local observables sensitive to the physics of extended objects must be exploited. A well-known example is the disorder operator in classical or quantum Ising models~\cite{KadanoffPRB1971, Fradkin2016}, which takes on a finite expectation value in the disordered phase. In the dual description, the disorder operator becomes the Wilson loop operator in a $\Z_2$ gauge theory~\cite{Wegner1971}, which is able to distinguish confined and deconfined phases. In a closely related line of development, generalized global symmetries, known as ``higher-form'' symmetries~\cite{Nussinov2006, Nussinov2009, Gaiotto_2015,ji2019categorical, kong2020algebraic}, have been introduced as a general theoretical framework to systematically organize non-local observables.  They offer new perspectives to quantum phases of matter that bridge the Landau-Ginzburg-Wilson paradigm of spontaneous symmetry breaking and more exotic phenomena of topological order.  

While extended observables (and the related higher-form symmetries) have already found numerous conceptual applications, more quantitative aspects, such as their scaling at quantum criticality above $(1+1)d$, are still not systematically understood. Recently, the Ising disorder operator, which serves as the order parameter of a $\Z_2$ 1-form symmetry, was computed by quantum Monte Carlo (QMC) simulation at the $(2+1)d$ Ising transition~\cite{JRZhao2020} and new universal scaling behavior was identified. It is important to understand the generality of these features in the broad context of quantum criticality, and the relation between the universal feature to intrinsic CFT data.

In this work we make progress towards answering these questions. We show that the logarithmic corner correction (to be defined below) to the disorder operator is generally present in U(1) CFTs in $(2+1)d$, and the universal coefficient can be related to the current central charge in the limit when the associated U(1) transformation is close to the identity. We then compare these results with unbiased QMC simulations of the disorder parameter across the superfluid-insulator transition in a Bose-Hubbard model, the prototypical example of continuous symmetry breaking transition. We find that as expected the disorder operator obeys the perimeter law in the insulating phase, and acquires a multiplicative logarithmic violation in the superfluid phase. At the critical point, we compute the corner correction and confirm the analytical predictions in the limit of small U(1) rotation angle. For more general CFTs, we derive the universal corner correction near the smooth corner limit, which is controlled by intrinsic defect CFT data. 

\begin{figure}[htp!]
	\centering
	\includegraphics[width=\columnwidth]{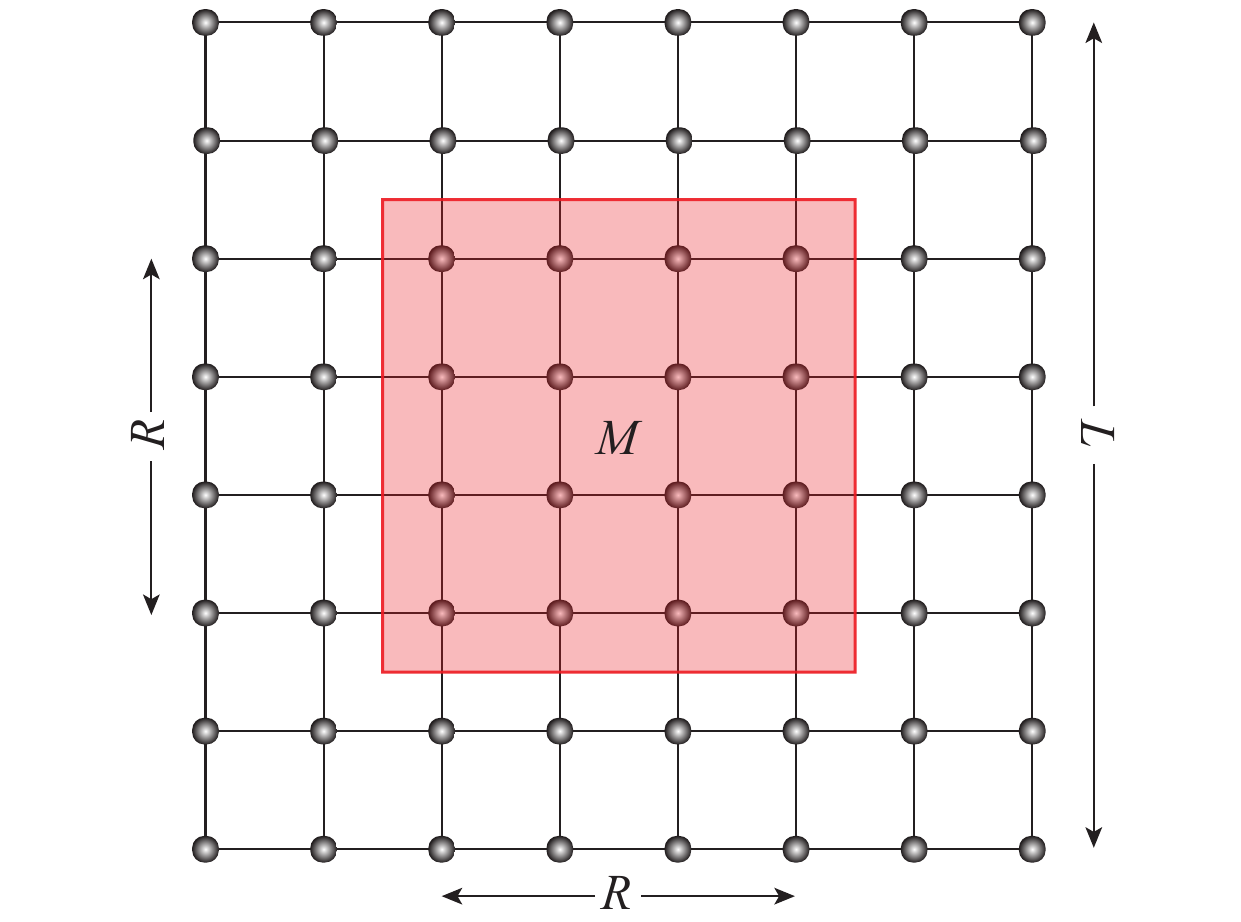}
	\caption{Disorder operator $X_M$ applied on regions with size $R \times R$ and perimeter $l=4R-4$ in the $L \times L$ square lattice of Boson-Hubbard model.}
	\label{fig:fig1}
\end{figure}

{\it{Disorder operator.-}}
Let us start from general considerations. For a $(2+1)d$ quantum lattice system with U(1) symmetry, we define a disorder operator in the following way: Suppose the U(1) symmetry transformations are implemented by $U(\theta)=\prod_\mb{r}e^{i\theta n_\mb{r}}$ where $n_\mb{r}$ is the charge on site $\mb{r}$. For a region $M$, we define 
\begin{equation}
	X_M(\theta)=\prod_{\mb{r}\in M}e^{i\theta n_\mb{r}}.
	\label{}
\end{equation}
The disorder parameter is the expectation value $\langle X_M(\theta)\rangle$ on the ground state. We note that the definition can be straightforwardly adapted to other symmetry group~\cite{ji2019categorical, JRZhao2020}.  

{\it{Scaling of the disorder parameter.-}} 
Next we discuss the scaling behavior of $X_M(\theta)$ in various phases of Bose-Hubbard model, especially the dependence on the geometry of $M$. In an insulating phase, $\langle X_M(\theta)\rangle$ is expected to obey a perimeter law $|\langle X_M(\theta)\rangle|\sim e^{- a_1(\theta) l}$, where $l$ is the perimeter of the region $M$. The perimeter dependence in this case can be absorbed into a local boundary term in the definition of the operator $X_M(\theta)$, and after the redefinition $|\langle X_M(\theta)\rangle|$ is finite for arbitrarily large $M$~\cite{HastingsPRB2005}. In the superfluid phase, on the other hand, it was found in \Refcite{Lake2018} that $|\langle X_M(\theta)\rangle|\sim e^{- b(\theta) l \ln l}$,  a weaker decay than the area law for a discrete symmetry breaking state, but still can not be remedied by any local counter-term on the boundary of $M$. In this sense, the disorder operator serves as an ``order parameter'' for the disordered (i.e. insulating) phase~\cite{Fradkin2016, Levin2019}.
	
We now focus on the disorder parameter in a quantum critical state described by a CFT at low energy. Previous studies of the  $(2+1)d$ Ising CFT and other gapless critical field theories~\cite{XCWu2020} suggest that $\ln |\langle X_M(\theta)\rangle|$ takes the following form for a rectangle region: 
\begin{equation}
	\ln |\langle X_M(\theta)\rangle|=-a_1 l+s \ln l + a_0.
\label{eq:eq3}
\end{equation}
Here the dependence on $\theta$ for the coefficients is suppressed.
The logarithmic correction, which translates into a power law $l^s$ in $|\langle X_M\rangle|$, originates from sharp corners of the region. In general $s$ is a universal function of both $\theta$ and the opening angle(s) of the corners (all $\pi/2$ in this case)~\footnote{Similar corner contributions were known to exist for Renyi entropy in a CFT, which can be understood as the disorder parameter of the replica symmetry}. We conjecture that the corner correction is a generic feature for disorder operators in any $(2+1)d$ CFT. Below we present new analytical arguments to support the conjecture and also connect the universal coefficient $s$ to intrinsic CFT data.

The first argument works for any CFT with global U(1) symmetry in the limit $\theta\rightarrow 0$. For small $\theta$, the Taylor expansion of $X_M(\theta)$ to $\theta^2$ order is given by
	\begin{equation}
		\langle X_M(\theta)\rangle\approx 1 - \frac{\theta^2}{2}\int_M \dd^2\mb{r}_1 \int_M\dd^2\mb{r}_2\, \langle n(\mb{r}_1)n(\mb{r}_2)\rangle.
		\label{eqn:expansion}
	\end{equation}
	Here $n(\mb{r})$ is the charge density in the continuum limit, and without loss of generality we assume $\langle n(\mb{r})\rangle=0$ so the first-order correction in the expansion vanishes. 
It is well-known that in a CFT with U(1) symmetry, the two-point function of the conserved charge density takes the following universal form:
\begin{equation}
	\langle n(\mb{r}_1)n(\mb{r}_2)\rangle = -\frac{C_J}{(4\pi)^2}|\mb{r}_1-\mb{r}_2|^{-4}.
	\label{}
\end{equation}
Here $C_J$ is the current central charge of the CFT, which is proportional to the universal DC conductivity $\sigma=\frac{\pi}{16}C_J$~\cite{Fisher1990}. We can now evaluate \eqnref{eqn:expansion} for $M$ a $L\times L$ square region. The integral has UV divergence, and once regularized we obtain
\begin{equation}
	\langle X_M(\theta)\rangle\approx 1- \frac{\theta^2 C_J}{(4\pi)^2}\left[\Big(1+\frac{\pi}{2}\Big)\frac{l}{8\delta}-  \ln \frac{l}{\delta}\right],
	\label{}
\end{equation}
where $\delta$ is a short-distance cutoff. Details of the evaluation of the integral in \eqnref{eqn:expansion} can be found in the Supplementary Material. Therefore we find 
\begin{equation}
	s(\theta)\approx \frac{C_J}{(4\pi)^2}\theta^2, \theta\rightarrow 0.
	\label{eq:eq7}
\end{equation}
One can also show that such logarithmic correction is absent when $M$ is a disk.

\begin{figure}[htp!]
	\centering
	\includegraphics[width=\columnwidth]{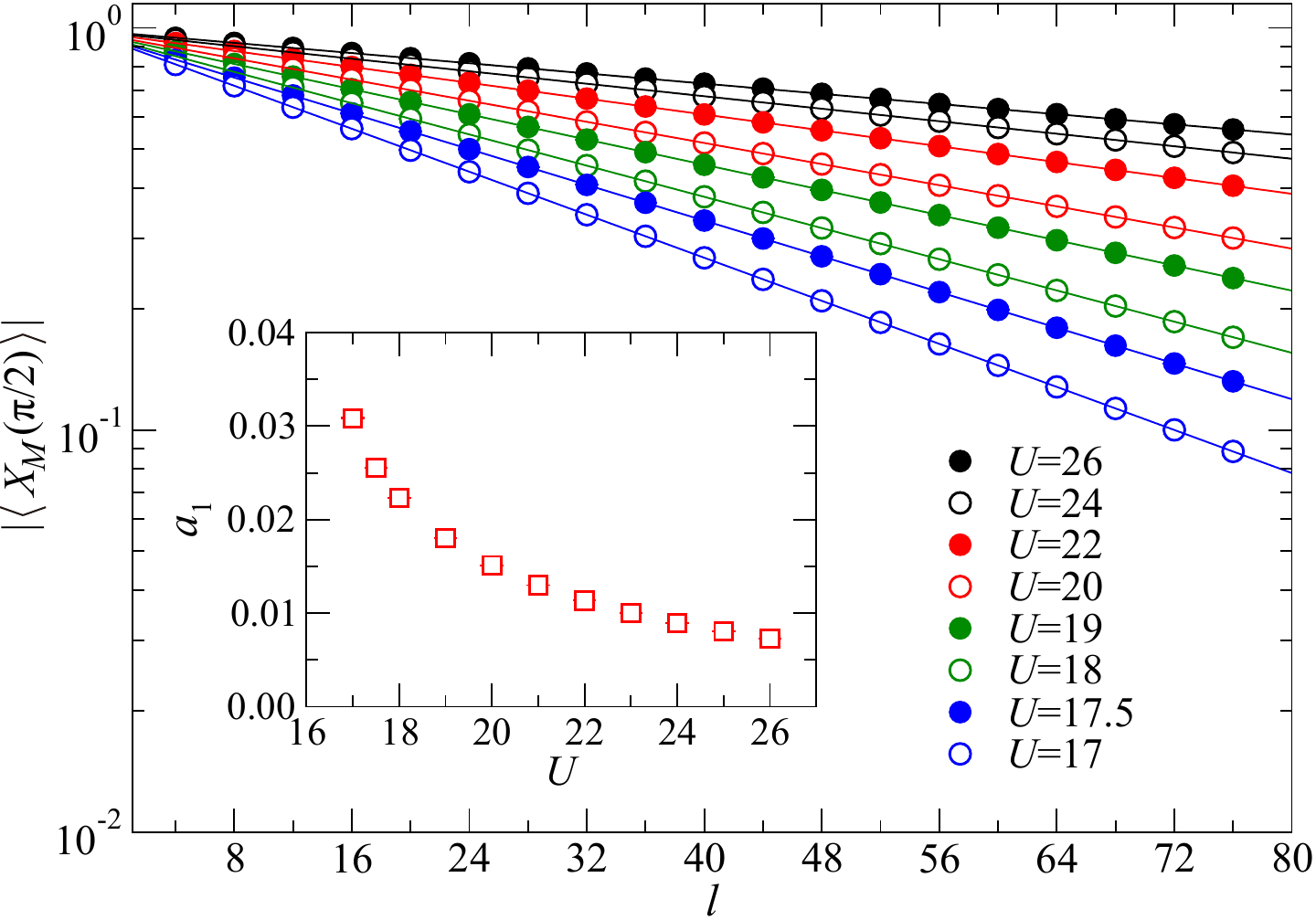}
	\caption{Disorder operator $|\langle X_M(\theta=\frac{\pi}{2}) \rangle|$ as function of $l$ in the MI phase (from $U=17$ to  $U=26$) with system sizes up to $L=40$ and $R\in [1,L/2]$. Dots are the QMC results with error bars smaller than the symbol size and the solid lines are the fitting with function $e^{-a_1 l}$. Inset shows the obtained $a_1$ as a function of $U$.}
	\label{fig:fig2}
\end{figure}

We now turn to disorder operators in a generic CFT. The universal coefficient $s$ is generally a function of the opening angle(s) of the corners of the region $M$. In the previous case of a square $M$, there are four corners with opening angle $\pi/2$. We now focus on the contribution from one corner, whose opening angle $\alpha$ is close to $\pi$ (so the corner is nearly smooth).  Under a generally accepted assumption about RG flow of defect lines in a CFT we have the following formula
\begin{equation}
	s(\alpha) = \frac{C_\mathrm{D}}{12}(\pi-\alpha)^2, \alpha\rightarrow \pi.
	\label{eqn:smooth}
\end{equation}
Here $C_\mathrm{D}$ is the defect central charge, a universal quantity for the disorder operator (see the supplementary material for the definition of $C_\mathrm{D}$ and the derivation of \eqnref{eqn:smooth})~\cite{Billo2013, Gaiotto2013}. A very similar relation was known for entanglement entropy in CFTs~\cite{Casini_2007, Bueno2015, Bueno_2015, Faulkner_2016, Bianchi_2016, WK2019}, and the derivations follow essentially the same idea. We stress the generality of \eqnref{eqn:smooth}, which holds for any disorder parameter in $(2+1)d$ CFTs.

While \eqnref{eq:eq7} and \eqnref{eqn:smooth} are valid only for small parameter regimes, they provide strong evidence that the logarithmic corner corrections are universally present. In the following section we perform a systematic study of $|\langle X_M(\theta) \rangle|$ in a U(1) boson lattice model with unbiased QMC computation and verify the analytical result \eqnref{eq:eq7}. We leave the lattice study of \eqnref{eqn:smooth} for future works.

\begin{figure}[htp!]
	\centering
	\includegraphics[width=\columnwidth]{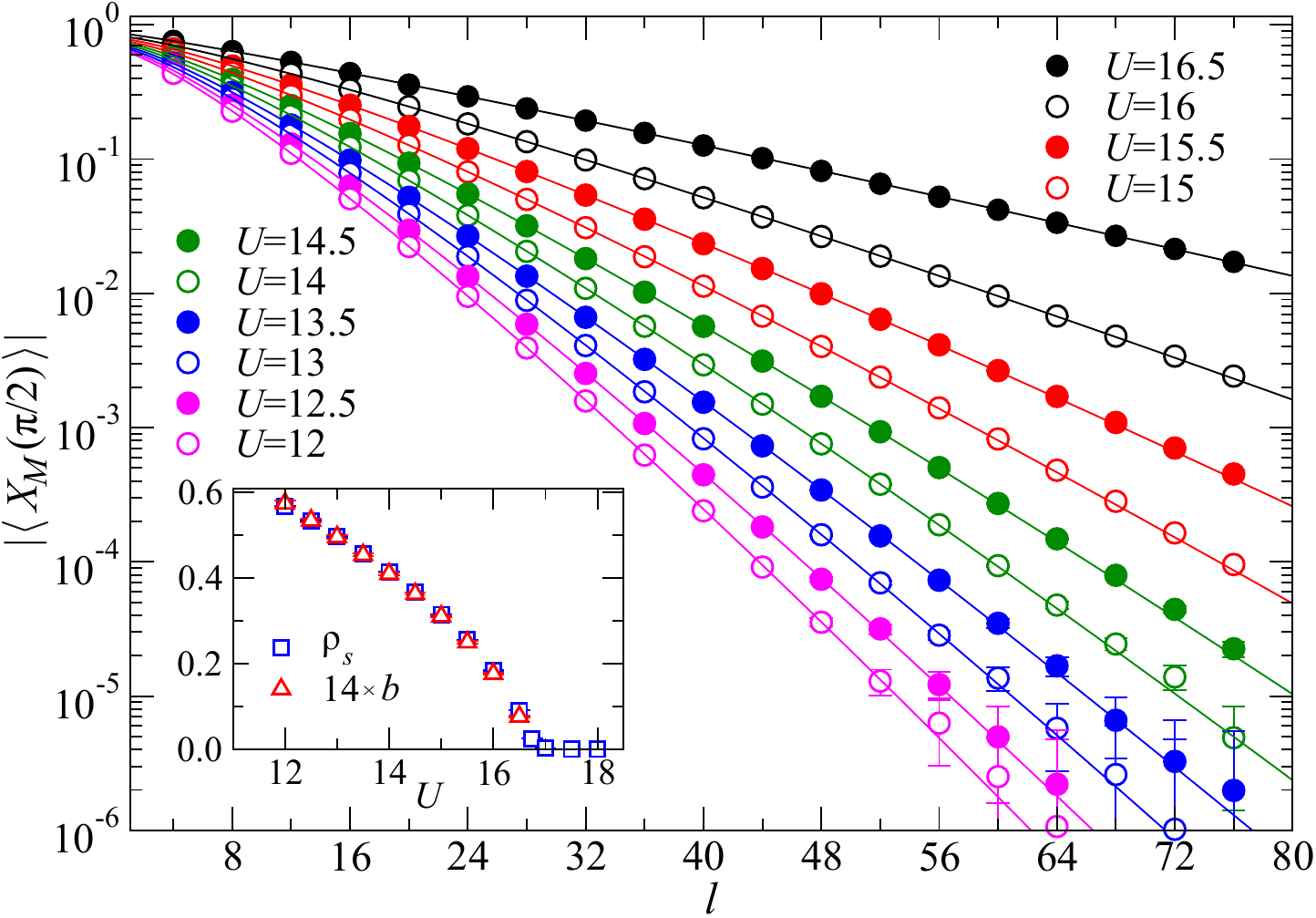}
	\caption{Disorder operator $|\langle X_{M}(\theta=\frac{\pi}{2}) \rangle|$ as function of $l$ in the SF phase (from $U=16.5$ to $U=12.0$) with system sizes up to $L=40$ and $R\in [1,L/2]$.  Dots are the QMC results with error bars smaller than the symbol size and the solid lines shows the fit with function $e^{-b l\ln\frac{l}{c}}$. Inset shows the fitting parameter $b$ as a function of $U$ as well as the superfluid stiffness $\rho_s$ for the same parameter sets.}
	\label{fig:fig3}
\end{figure}

{\it{Superfluid-insulator transition.-}}
While the field-theoretical approach has yielded general results about universal features of the disorder parameter, one has to take various limits, e.g. $\theta\rightarrow 0$, to make progress analytically. To probe parameter regimes where analytical results are not available, we turn to numerical simulations.

We consider the Bose-Hubbard model on the square lattice, which provides a concrete realization of the superfluid-insulator transition~\cite{Fisher1989b}. The Hamiltonian takes the following standard form:
\begin{equation} 
	H=-t \sum_{\langle ij\rangle}\left(b_i^\dag b_j + \text{h.c.}\right) + \frac{U}{2}\sum_{i}n_i(n_i-1) - \mu\sum_{i}n_i,
\label{eq:eq1}
\end{equation} 
where $b^\dag (b)$ is the boson creation (annihilation) operator,  $n_i=b_i^\dag b_i$ the boson number, $t>0$ is the hopping between nearest-neighbor sites on the square lattice,  $U>0$ is the on-site repulsion, and $\mu$ is the chemical potential. We set $t=1$ as the unit of energy for convenience.

The ground state of this model has two phases: a Mott insulator (MI) phase for large $U/t$ and a superfluid (SF) phase for small $U/t$, separated by a continuous phase transition. At integer filling, the transition belongs to the 3D XY universality class, also known as the O(2) Wilson-Fisher theory~\cite{Fisher1989b}. For $\langle n \rangle=1$, the critical point is located at $U_c/t=16.7424(1)$, $\mu/t=6.21(1)$, determined from previous works~\cite{Capogrosso-SansonePRA2008,SoylerPRL2011,ChenPRL2014}.

\begin{figure}[htp!]
	\centering
	\includegraphics[width=\columnwidth]{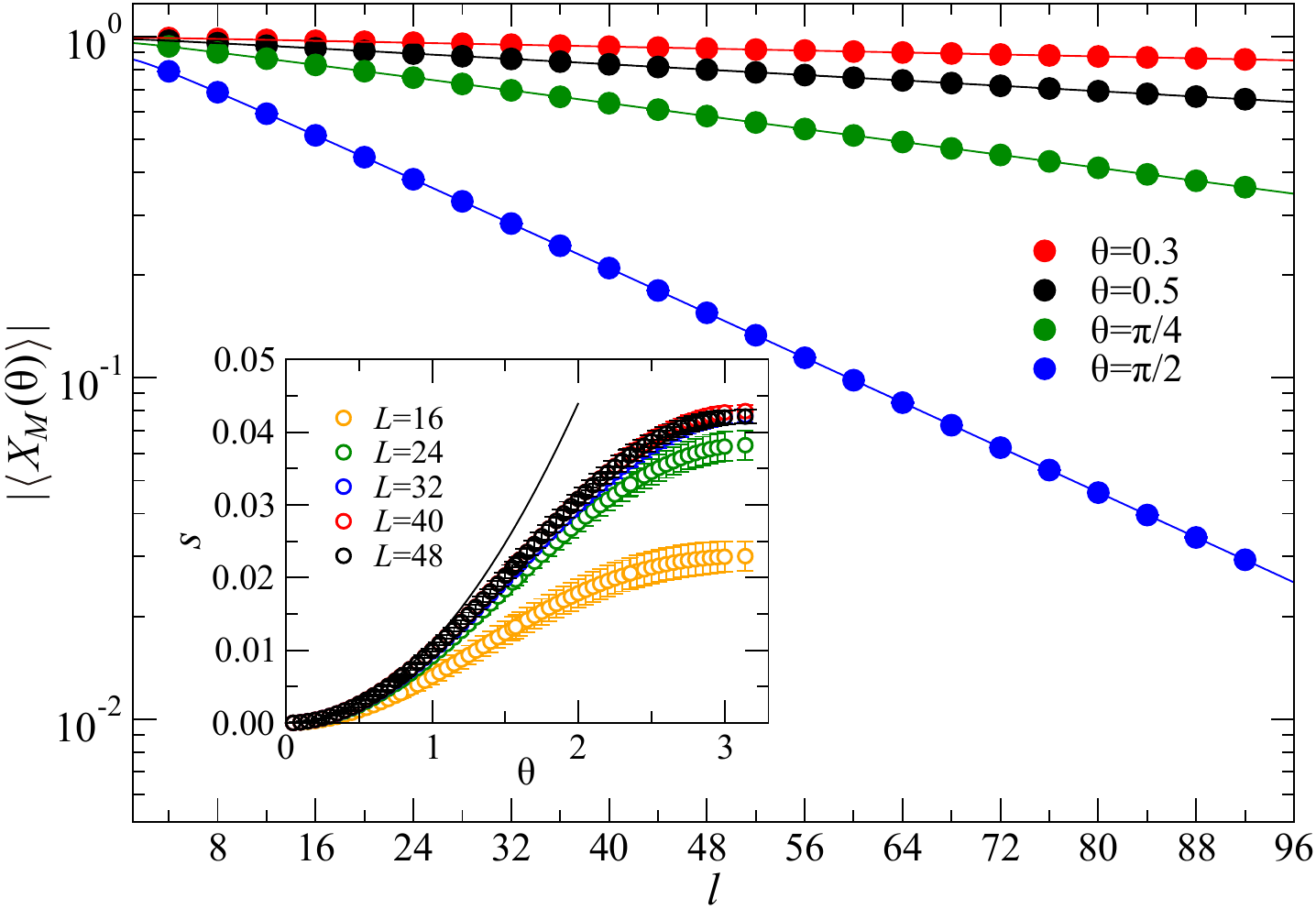}
	\caption{Disorder parameter $|\langle X_M(\theta) \rangle|$ as function of $l$ at the critical point  ($U_c=16.7424$) with $\theta=0.3, 0.5, \pi/4, \pi/2$ and system sizes up to $L=48$, $R\in [1,L/2]$. Dots are the QMC results with error bars smaller than the symbol size and the solid lines show the fit by the function in Eq.~\eqref{eq:eq3}. Inset shows the $\theta$ dependence of universal coefficient $s$ for different system sizes. For small $\theta$, a quadratic dependence clearly manifests. The finite-size results converge for $L=40, 48$ and fitting with Eq.~\eqref{eq:eq7} yields the coefficient $\frac{s}{\theta^2}\approx 0.011(1)$, close to the exact value $\frac{C_J}{(4\pi)^2}=0.01145$.}
	\label{fig:fig4}
\end{figure}

{\it{Numerical Results.-}}
We choose the region $M$ to be a $R\times R$ square region in the lattice, with perimeter $l=4R-4$. For an illustration see Fig.~\ref{fig:fig1}.  To calculate the disorder operator of the Bose-Hubbard model, we employ large-scale stochastic series expansion QMC simulations~\cite{Syljuaasen2002,Sandvik2010,ZYMeng2008}, and compute the expectation value of $X_M(\theta)$ on finite lattice with $L=\beta=1/T$ and $R\in[1,L/2]$ to access the thermodynamic limit.
  For the MI and SF phases we fix $\theta=\frac{\pi}{2}$. 

First, in the MI phase the disorder parameter decays according to the perimeter law. This is shown in Fig.~\ref{fig:fig2}, where in a semi-log plot, the relation $|\langle X_M(\frac{\pi}{2})\rangle| \sim e^{-a_1 l}$ is clearly seen. We also observe that the coefficient $a_1$ decreases monotonically with $U$ from $U=17$ to $26$, as shown in the inset of Fig.~\ref{fig:fig2}, consistent with the theoretical expectation.

Inside the SF phase, the disorder parameter decays more rapidly with the perimeter $l$, as depicted in Fig.~\ref{fig:fig3}.  We find that the data in Fig.~\ref{fig:fig3} can be well fitted by the function $|\langle X_{M}(\frac{\pi}{2})\rangle |\sim e^{-b l\ln{\frac{l}{c}}}$. Interestingly, the coefficient $b$ extracted from the fit is proportional to the superfluid stiffness $\rho_s=\langle W_{\mathbf{x}}^2+W_{\mathbf{y}}^2 \rangle /(4\beta t)$ (where $W_{\mathbf{x,y}}$ is the winding number along $\mathbf{x}$ or $\mathbf{y}$ direction), inside the SF phase (shown in the inset of Fig. \ref{fig:fig3}), also consistent with theoretical analysis~\cite{Lake2018}.

\begin{figure*}[htp!]
	\centering
	\includegraphics[width=\textwidth]{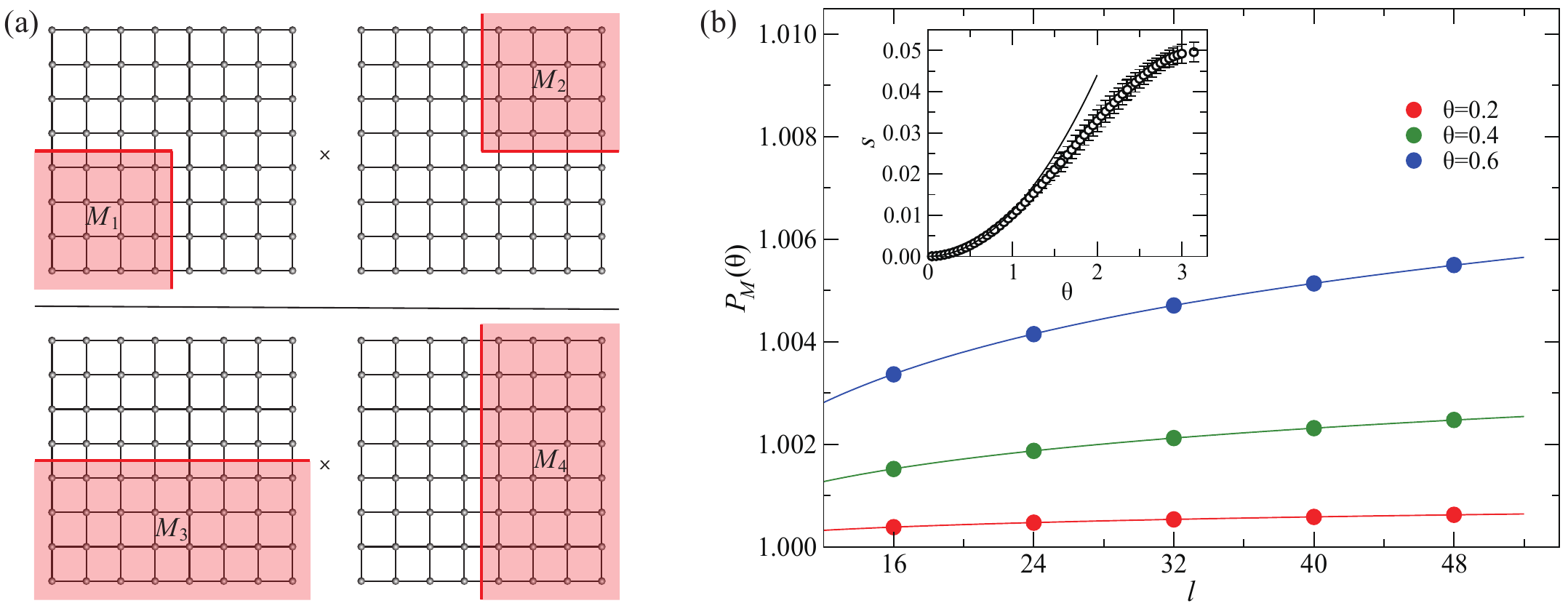}
	\caption{(a) Regions $M_1$, $M_2$, $M_3$, and $M_4$ used to determine the corner contribution in Eq.~\eqref{eq:eq3}  on the  square lattice with open boundary condition, and the perimeter of $M_1\cup M_2$ is equal to that of $M_3\cup M_4$. (b) $P_M(\theta)$ as function of $l(=L)$ at the critical point ($U_c=16.7424$) with the system size $L=16, 20, 24, 28, 32,36, 40, 44, 48$. The data points are fitted by $l^{s(\theta)/2}$ for each $\theta$, as shown by the solid lines in the main panel, to obtain $s(\theta)$ for $\theta\in [0,\pi]$ denoted by the black dots in the inset. Then fitting $s(\theta)$ with $\frac{C_J}{(4\pi)^2} \theta^2$ for $\theta \in [0,0.3]$, one finds $\frac{C_J}{(4\pi)^2} \approx 0.0109(5)$. The solid line in the inset shows the fit.}
	\label{fig:fig5}
\end{figure*}

We now turn to the critical point $U=U_c=16.7424$. Numerical results of $\langle X_M(\theta)\rangle$ as a function of $l$ for various values of $\theta$ are shown in Fig.~\ref{fig:fig4}, and we find that the data can be fitted with the scaling form in Eq.~\eqref{eq:eq3} with good quality, in that the coefficient of the sub-leading logarithmic term $s$, as shown in the inset of Fig.~\ref{fig:fig4}, clearly manifests a quadratic dependence with respective to $\theta$, when $\theta$ is small. However, one might worry whether such fitting can reliably extract the coefficient $s$ of the sub-leading logarithmic term, as the perimeter contribution clearly dominates. 

We thus apply a different method adapted from \Refcite{KallinJS2014} to directly extract the corner correction. In this approach, we work on a $L\times L$ square lattice with open boundaries. We measure disorder parameters for each of the four regions $M_{1,2,3,4}$ as shown in Fig.~\ref{fig:fig5}(a). The regions are chosen such that the perimeter of $M_1\cup M_2$ is equal to that of $M_3\cup M_4$. So the following combination 
\begin{equation}
	P_M(\theta)=\left|\frac{\langle X_{M_1}(\theta) \rangle \langle X_{M_2}(\theta) \rangle} {\langle X_{M_3}(\theta) \rangle \langle X_{M_4}(\theta) \rangle}\right|
	\label{eq:eq10}
\end{equation}
cancels the leading term  $a_1l$ in \eqnref{eq:eq3}. Since both $M_1$ and $M_2$ contain one $\pi/2$ corner, we expect $P_M(\theta)\sim l^{s(\theta)/2}$, which can then be used to determine $s(\theta)$.  We find that the two methods give basically identical values of $s$ for small $\theta$, although there are small discrepancies when $\theta$ gets close to $\pi$. The full function $s(\theta)$ for $\theta\in[0,\pi]$ determined from the latter method is shown in the inset of Fig. \ref{fig:fig5}(b), which is very close to the function: $0.047 \sin^2(\frac{\theta}{2})$.

To corroborate the analytical results, we examine more closely the function $s(\theta)$ as $\theta\rightarrow 0$. As shown in the insets of Figs. \ref{fig:fig4} and \ref{fig:fig5}, $s(\theta)$ exhibits a clear $\theta^2$ dependence, and the coefficient is found to be $0.011(1)$ for the direct fitting method (Fig. \ref{fig:fig4}) and $0.0109(5)$ for the second method (Fig. \ref{fig:fig5}). Using the formula \eqnref{eq:eq7} and the best estimate $C_J=1.8088$ for the O(2) Wilson-Fisher CFT~\cite{ Krempa2014, Katz2014, ChenPRL2014, Chester_2020}, we obtain the theoretical value for the proportionality constant $\frac{C_J}{(4\pi)^2}=0.01145$. The numerical results agree quite well with the theory.

{\it{Discussions.-}}
We briefly discuss future directions. An immediate question is to verify the smooth corner limit Eq.~\eqref{eqn:smooth} in a lattice model, which would provide a way to extract the defect central charge. We have mainly considered the modulus of the disorder parameter $\langle X_M\rangle$. An interesting question is to understand the phase of $\langle X_M\rangle$ and how it depends on intrinsic CFT data. According to the small $\theta$ expansion, the leading imaginary part appears at $\theta^3$ order, which is then related to the three-point function of density operator. 

In summary, { we develop a new computational and theoretical toolkit about non-local observables -- the disorder operator -- and the associated higher-form symmetry in lattice model of quantum many-body systems, and demonstrate that it can directly reveal the CFT data of the critical point beyond the conventional local observables. This offers a new concept and technique in understanding new aspect of phase transitions.} It would be interesting to study other conformal field theories, such as O($n$) symmetry-breaking transitions~\cite{Lohoefer2015,NSMa2018} and even more unconventional phase transitions such as the deconfined quantum critical points~\cite{Senthil2004,YQQin2017,CWang2017,YCWang2021}, or non-conformal scale-invariant theories such as the Lifshitz critical point~\cite{FradkinPRL2006}.

{\it{Acknowledgement.-}}
We would like to thank William Witczak-Krempa and Shu-Heng Shao for stimulating discussions which benefit the present work, as well as comments on the first draft. We are grateful for Chao-Ming Jian and Cenke Xu for correspondence and sharing unpublished work.  Y.C.W. acknowledges the supports from the NSFC under Grant No.~11804383 and No.~11975024, the NSF of Jiangsu Province under Grant No.~BK20180637, and the Fundamental Research Funds for the Central Universities under Grant No. 2018QNA39.  M.C. acknowledges support from NSF under award number DMR-1846109 and the Alfred P.~Sloan foundation. Z.Y.M. acknowledges support from the RGC of Hong Kong SAR of China (Grant Nos. 17303019, 17301420 and AoE/P-701/20), MOST through the National Key Research and Development Program (Grant No. 2016YFA0300502) and the Strategic Priority Research Program of the Chinese Academy of Sciences (Grant No. XDB33000000). We thank the Computational Initiative at the Faculty of Science and the Information Technology Services at the University of Hong Kong and the Tianhe platforms at the National Supercomputer Centers in Tianjin and Guangzhou for their technical support and generous allocation of CPU time.

{\it{Note added.-}} We would like to draw the reader’s attention to a closely related work by X.-C. Wu, C.-M. Jian and C. Xu~\cite{XCWu2021} in the same arXiv listing. We also become aware of an upcoming work by B. Estienne, J.-M. St\'ephan and W. Witczak-Krempa on related topics~\cite{Estienne2021}.

\bibliographystyle{apsrev4-1}
\bibliography{u1hfsb}

\setcounter{equation}{0}
\renewcommand{\theequation}{S\arabic{equation}}
\renewcommand{\thefigure}{S\arabic{figure}}

\newpage

\begin{widetext}
\section*{Supplemental Material}
	
\subsection{Evaluation of the integral in Eq. \eqref{eqn:expansion}}
We evaluate the integral
\begin{equation}
	I=\int_{0}^{R}\dd x_1\dd x_2\dd y_1\dd y_2\, \frac{1}{[(x_1-x_2)^2+(y_1-y_2)^2]^2}.
	\label{}
\end{equation}
First integrate over $x_1$ and $x_2$ to obtain
\begin{equation}
	I=\int_0^{R}\dd y_1\dd y_2\, \frac{R}{(y_1-y_2)^3}\arctan \frac{R}{y_1-y_2}.
	\label{}
\end{equation}
Then with a change of variables $y_+ = y_1+y_2, y_-=y_2-y_1$, $I$ can be rewritten as
\begin{equation}
	I=2R\int_0^R \dd y_+\,\int_{\delta}^{y_+} \dd y_-\,\frac{1}{y_-^3}\arctan \frac{R}{y_-}.
	\label{}
\end{equation}
Here $\delta$ is a short-distance cutoff. Evaluate the integral over $y_-$ and Taylor expand in powers of $\delta/R$,
\begin{equation}
	\begin{split}
	I&=\int_\delta^R \dd y_+\,\left(\frac{1}{y_+}+\frac{1}{R}\arctan \frac{y_+}{R}-\frac{R}{y_+^2}\arctan\frac{R}{y_+}\right)+\frac{\pi R^2}{2\delta^2}-\frac{R}{\delta}\\
	&= \frac{\pi R^2}{2\delta^2}-\left( 1+\frac{\pi}{2} \right)\frac{R}{\delta} + 2\ln\frac{R}{\delta}+\frac{\pi}{2}-\ln 2 
	\end{split}
	\label{}
\end{equation}
The first term $\frac{\pi R^2}{2\delta^2}$ is the UV divergence, which can be cancelled by adding a local counterterm $\frac{\pi}{2\delta^2}\delta(\mb{r}_1-\mb{r}_2)$ in $\langle n(\mb{r}_1)n(\mb{r}_2)\rangle$. The remaining terms give the result quoted in the main text.

\subsection{Corner contribution in the smooth limit}

In this section we present a field-theoretical derivation of the corner contribution in the smooth limit. Our derivation closely follows \Refcite{Bianchi_2016} and \Refcite{Faulkner_2016} for corner correction to entanglement entropy in CFTs (see also \Refcite{WK2019}).

Throughout the section we will work with Euclidean formalism. The disorder operator inserts a symmetry twist defect loop into the Euclidean path integral of the field theory, and the ground state expectation value is given by the partition function in the presence of such a defect loop. We will assume that the symmetry twist defect line flows to a Conformal Defect Line (CDL) at low energy, which is widely believed to be true in any CFT. 

To be more concrete, let us fix a straight defect line $\mathcal{C}$ at $x_2=x_3=0$. We denote the corresponding disorder operator by $X_\mathcal{C}$. Because of the defect line, the space-time symmetry of the $3d$ CFT, or more precisely the connected component, is reduced from the conformal group SO$(4,1)$ to $\mathbb{R}\times \mathrm{SO}(2)$ where $\mathbb{R}$ is the translation symmetry along the defect line and $\mathrm{SO}(2)$ the residual rotation around the line. If at low energy the space-time symmetry is enlarged to $\mathrm{SO}(2,1)\times \mathrm{SO}(2)$, the defect line is said to be conformal.  For a more systematic account on defects in CFTs we refer the reader to \Refcite{Billo2016}.

A CDL supports a displacement operator $D^i, i=1,2$, which generates infinitesimal local deformations of the defect line in the transverse directions. It should be clear that $D^i$ transforms as a vector under the SO(2) group. The breaking of the translation symmetry in the transverse directions at the defect line leads to the following Ward identity:
\begin{equation}
	\partial_\mu T^{\mu i}(x) = \delta_\mathcal{C}(x) D^i(x).
	\label{}
\end{equation}
Here $\delta_{\mathcal{C}}$ denotes the delta function in the transverse space with support on the $\mathcal{C}$. The Ward identity essentially fixes the normalization of the displacement operator $D^i$, and imply the following universal two-point function:
\begin{equation}
	\langle D^i(x)D^j(y)\rangle_\mathcal{C}=\frac{C_\mathrm{D} \delta_{ij}}{|x-y|^4}.
	\label{eqn:2pt_D}
\end{equation}
Here the expectation value is defined as
\begin{equation}
	\langle O\rangle_\mathcal{C}\equiv\frac{\langle X_\mathcal{C} O\rangle}{\langle X_\mathcal{C}\rangle}.
	\label{}
\end{equation}
The universal coefficient $C_\mathrm{D}$ is an intrinsic property of the CDL, which we call the defect central charge. $C_\mathrm{D}$ should be uniquely fixed by the type of the defect line. In the case of a symmetry twist defect, the defect type is determined by the symmetry transformation.

Using the Ward identity, we can calculate the change of the disorder parameter due to small deformation $\xi^i(x)$ of the defect line by integrating $D^i$:
\begin{equation}
	\frac{\langle X_{\mathcal{C}'}\rangle}{\langle X_{\mathcal{C}}\rangle}= \left\langle \exp\left(-\int_{\mathcal{C}} \xi^i(x)D^i(x)\right)\right\rangle_\mathcal{C}=1+\frac{1}{2}\int_{\mathcal{C}}\dd x\int_{\mathcal{C}} \dd y\, \langle D^i(x)D^j(y)\rangle_\mathcal{C}\,\xi^i(x)\xi^j(y) + \cdots
	\label{eqn:deform}
\end{equation}
Here $\mathcal{C}'$ denotes the deformed defect line.
Notice that the first-order term should vanish by the assumption that the defect line is conformal.

To set up the computation of the corner correction,  we deform the defect line in the interval $x_1\in [-L,L]$ by the following shape parameter~\cite{Faulkner_2016}:
\begin{equation}
	\xi^2(x)=\begin{cases}
		\frac{\eta}{2L}(L^2-x_1^2) & |x_1|<L\\
		0 & |x_1|\geq L
	\end{cases}, \xi^3(x)=0.
	\label{}
\end{equation}
This way two corners at $x_1=\pm L$ are introduced, both with opening angle $\alpha=\pi-\eta$.  According to \eqnref{eqn:deform}, we have
\begin{equation}
	\begin{split}
		\frac{\langle X_{\mathcal{C}'}\rangle}{\langle X_{\mathcal{C}}\rangle} &\approx 1+\frac{C_\mathrm{D}}{2}\int\dd x_1\dd x_2\, \frac{\xi(x_1)\xi(x_2)}{(x_1-x_2)^4}\\
&=1+\frac{\alpha^2 C_\mathrm{D}}{8L^2}\int_{-L}^L\dd x_1\,\int_{-L}^L\dd x_2\, \frac{(L^2-x_1^2)(L^2-x_2^2)}{(x_1-x_2)^4}
	\end{split}
	\label{}
\end{equation}
The last integral contains UV divergence, which can be dealt with e.g. dimensional regularization~\cite{Faulkner_2016} or just introducing a short-distance cutoff. The final result is given by
\begin{equation}
	\frac{\langle X_{\mathcal{C}'}\rangle}{\langle X_{\mathcal{C}}\rangle} \approx 1+\frac{C_\mathrm{D}\eta^2}{6}\left(\ln \frac{L}{\delta}+\cdots\right),
	\label{}
\end{equation}
where all non-universal terms have been omitted.
Since there are two corners with the same opening angles, for one corner we obtain
\begin{equation}
	s(\alpha)=\frac{C_\mathrm{D}(\pi-\alpha)^2}{12}.
	\label{eqn:s}
\end{equation}

We also notice that if we simply take the expansion \eqnref{eqn:deform} as a starting point, then the scale invariance alone suffices to fix the ``kernel'' $\langle D(x)D(y)\rangle_\mathcal{C}$ to the universal form \eqnref{eqn:2pt_D}, and the rest follows (see \Refcite{WK2019} for a similar argument for entanglement entropy). Therefore \eqnref{eqn:s} should hold in more general scale-invariant theories.

\end{widetext}

\end{document}